# Single Channel Cutaneous Electrogastrography on Local White Rabbit (*Oryctolagus cuniculus*) and Its Electrogastrogram Classification Algorithm

## Elektrogastrografi Kutan Satu Channel pada Kelinci Putih Lokal (Oryctolagus cuniculus) Beserta Algoritma Pengelompokan Elektrogastrogramnya

**Tyas Pandu Fiantoro[1], Lussya Eveline Rawar[2]**

[1]Electronic Engineering undergraduate, Universitas Gadjah Mada
[2]Veterinary Sciences undergraduate, Universitas Gadjah Mada
Email: eltroyaz@mail.ugm.ac.id

### Abstract

It has been a common practice to place electrodes based on external landmarks, rather than locating the appropriate organ first by imaging techniques such as CT scan, ultrasonography, etc. Therefore, aside from abiding the cutaneous EGG (electrogastrography) electrodes placement rule, identification of its waveform should be performed to ease the validation of one's EGG recording method. This research focused on the assembly of EGG instrument, its performance testing, and its usage on 13 local white rabbits (*O. cuniculus*). A total of 72 recordings obtained and processed. Data processing implies EGG parameterisation based on segmentation and time domain analysis. Therefore this research gives an insight of an EGG recording method that could be applied on another preclinical, veterinary, and even for clinical examination.

### Abstrak

Sudah menjadi kebiasaan umum bahwa penempatan elektroda dilaksanakan sebatas hapalan yang mengacu pada posisi/anatomi luarnya saja, tanpa melakukan penentuan letak organ yang sebenarnya melalui teknik pencitraan semacam USG (*ultrasonography*), CT (*computed tomography*), dan sejenisnya. Oleh karenanya, pengenalan bentuk gelombang EGG (elektrogastrografi) yang muncul akibat suatu tatacara pemasangan elektroda dapat dijadikan salah satu tolak ukur akan keabsahan tatacara EGG yang dilakukan. Penelitian ini dimulai dari pembuatan beserta pengujian instrumen EGG, dan kemudian teknik perekaman EGG dengan instrumen tersebut diterapkan pada 13 ekor kelinci putih lokal (*O. cuniculus*). Sebanyak 72 buah rekaman EGG berhasil diperoleh, dan diolah. Pengolahan data merujuk pada parameterisasi rekaman EGG berdasar segmentasi dan analisis domain waktu. Dengan demikian, apabila instrumen EGG dapat dibuat, diujikan, diterapkan, dan data EGG yang diperoleh dapat diolah, muncul sebuah tatacara yang dapat digunakan untuk keperluan pengujian pra-klinis, veteriner, hingga pengujian klinis pada manusia.

**Kata kunci:** *Oryctolagus cuniculus*, elektrogastrografi, letak elektroda, algoritma, klasifikasi pola

## Introduction

EGG commonly used for diagnosing tachygastric, normogastric, and bradygastric condition. EGG segmentation and its waveform classification is uncommon compared to ECG segmentation and waveform classification that already been standardised worldwide.

This research focused on the assembly of EGG instrument, its performance testing, and its usage on 13 local white rabbits (*O. cuniculus*). A standard depolarisation – plateau – repolarisation – resting waveform segmentation follows EGG acquisition. The electrogastrograph consists of an instrumentation amplifier connected with a single channel probe to a digital oscilloscope.

## Materials

The electrogastrograph assembled from a pair of conductive stud buttons as cutaneous electrodes (Prym Press Stud 555, Germany) that connected to an instrumentation amplifier unit based on AD620 chip (Analog Devices), copper "jumper" wires, 1 kΩ resistor with 1% tolerance, circuit board, and 7 cm x 5 cm plastic case (Lion Star). The output of the instrumentation amplifier unit is fed up into two channels digital oscilloscope (GWINSTEK GDS 1102-A-U). Recorded EGG stored at a 16 GB flash disk as comma separated value (CSV) files.

Additional tools used are a soldering device (Goot, type KX-30R 220V/30W) with copper tip type KX-20~100, a digital multimeter (Heles type UX 839-TR), a fur clipper (WAHL), scissors, tweezers, gloves, bandages, and colourful markers. Electrogastrograph response testing performed with a signal generator (Thurlby – Thandar) for generating the sinusoidal test input signal.

## Electrogastrograph Technical Details

Electrogastrograph instrument consisted of a pair cutaneous electrode, an instrumentation amplifier unit, and a digital oscilloscope. The instrumentation amplifier unit based on AD-620 IC chip with 8 pins. Pin 2 connected to the green electrode (to be placed cutaneously on a point between *pyloric sphincter – terminal anthrum*).

Pin 3 connected to the red electrode (to be placed cutaneously on a point between *esophageal sphincter – orad corpus*). There was a $R_G$ with 1000 Ω resistance and 1% tolerance, connecting pin 1 and pin 8. Pin 5 joined to a junction between two 9 volts battery, that also used as the reference point (ground) for the oscilloscope. Pin 4 attached to the positive pole of the first 9 V battery, and pin 4 attached to the negative pole of the second 9 V battery, in which the two batteries joined each other with the negative pole of the first battery attached to the positive pole of the second battery.

The output of the AD-620 chip (pin 6) fed into an input channel of digital oscilloscope GWINSTEK GDS 1102-A-U. Used sample rate was 10 Hz and recording duration was 400 seconds. This recording method covered the important EGG frequency range, that is 0.05 Hz to 4 Hz (Naruse *et al.*, 2000; Van der Voort, 2003; Kenneth and Robert, 2004; Sagami and Hongo, 2007; Matsuura, 2009). Any kind of filter was not used to prevent signal artifact. The schematic of the electrogastrograph is illustrated in Figure 1.

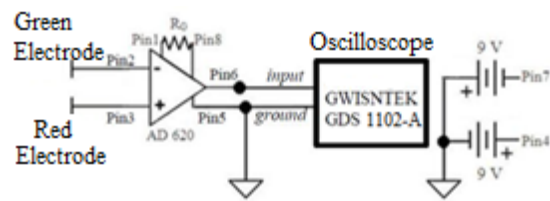

Figure 1. Electrogastrograph Schematic

## Electrogastrograph Performance Testing

The electrogastrograph response was tested using sinusoid signals with frequency of 0.01 Hz, 0.05 Hz, 0.5 Hz, and 0.1 Hz with 20 mV amplitude from a signal generator (Thurlby – Thundar). This frequency range was relevant to the EGG frequency range (Kaneoke *et al.*, 1992; Homma *et al*., 1998; Naruse *et al.*, 2000; Kenneth and Robert, 2004; Sagami and Hongo, 2007; Matsuura *et al.*, 2009). The response testing performed by feeding the generated input signal into channel 1 of the oscilloscope, together with feeding the resulted output from AD-620 unit into channel 2 of the oscilloscope. Both data then saved into a flash disk. Decibel scaled signal to noise ratio (*SNRdB*) then measured from the collected data. The electrogastrograph used in this research has a value of *SNRdB* ranging from 39 dB to 55 dB for frequency range of 0.05 Hz to 0.1 Hz.

### Electrode Placement

All rabbits' fur on the specific lumbar area (that would be the place for the electrodes) were clipped. The clipping area can be seen in Figure 2. The remaining fur aftershave is around 2 mm, as shown in Figure 3.

The placement of the two electrodes shown in Figure 4, 5, and 6, and agrees with Kaneoke and Sagami EGG method (Sagami *et al.*, 2007; Květina *et al.*, 2010), even with *Handbook of Electrogastrography* (Kenneth and Robert, 2004) but only without the reference electrode that placed on the minor curvature representative lumbar area. This recording scheme is using only one recording channel, hence this method classified as single channel cutaneous EGG.

The red electrode placed on the lumbar area that represents a point between *esophageal sphincter* and *orad corpus*. The green electrode placed on the lumbar area that represents a point between *pyloric sphincter* and *terminal anthrum*.

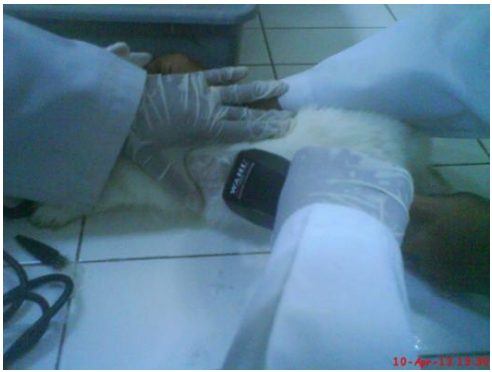

Figure 2. Shaving area

Before the electrodes were attached, a conductive gel was given on the respective lumbar area. To make sure the attached electrodes did not detach, a bandage is used as seen in Figure 7. This method did not hurt the experiential rabbits (*O. cuniculus*).

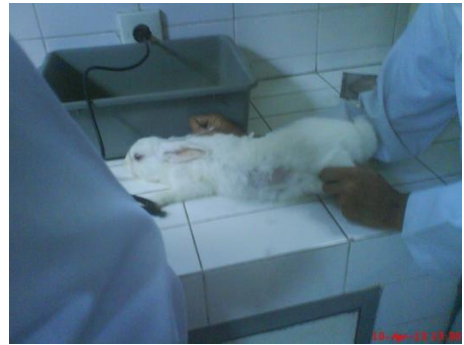

Figure 3. The clipped area

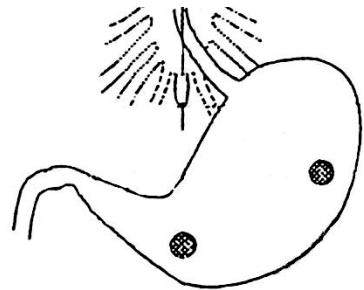

Figure 4. Two location of the electrodes

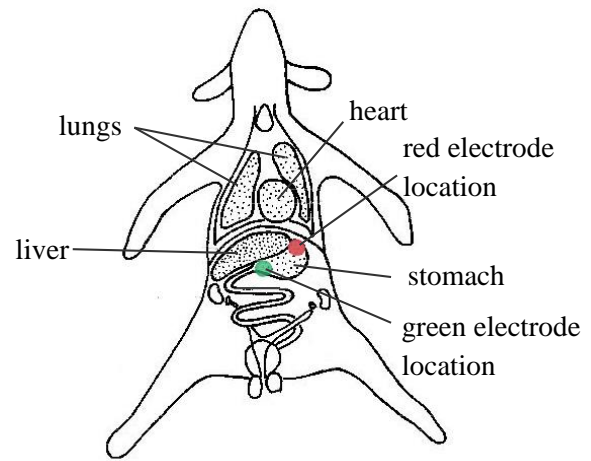

Figure 5. The location of red and green electrode

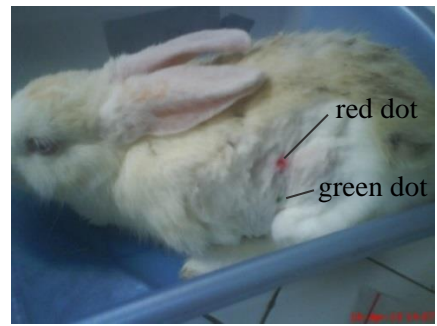

Figure 6. The red dot and the green dot

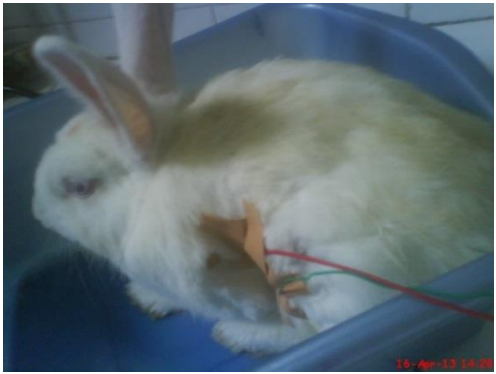

Figure 7. Bandage attached electrodes

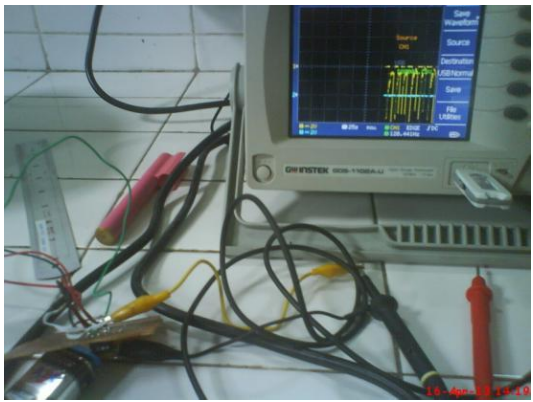

Figure 8. The electrogastrograph and the attached white flash disk.

Each red and green electrode connected to the input pins of AD-620 (pin 3 and pin 2 respectively). The output of AD-620 amplifier unit fed into digital oscilloscope (GWINSTEK GDS 1102-A-U). The recording performed in 400 seconds with a sampling rate of 10 Hz. Hence the maximum oscilloscope's storage memory of 4000 samples used up. This considered sufficient (Jieyun Yin, 2013) because compared to *Homo sapiens*, *Oryctolagus cuniculus* has a gastric wave rate 3 to 4 times faster. Saved EGG recording the transferred into a flash disk as seen in Figure 8.

## Electrogastrogram Segmentation

Obtained EGG then processed using SCILAB 5.1 for segmentation. EGG segments such as depolarisation, plateau, repolarisation, and resting potential segment were determined in this step. Amplitude comparison scheme was used for the resting potential segment classification. Peak detection scheme was used for depolarisation and repolarisation segment classification. The rest of the wave region that did not fit into previous classifications belongs to the plateau segment.

The amplitude comparison scheme performed by comparing the absolute amplitude value of the wave samples with a threshold potential called the activation potential $V_a$. This activation potential value could be tailored according to the peak amplitude of one's EGG profile. Usually, a value of 2 mV was used, but for certain cases where the peak amplitude was too weak, a value of 0.5 mV was used. Any wave sample with amplitude lower than $V_a$ considered to be a constituent of the resting potential segment.

Occasionally, there are wave samples with amplitude lower than $V_a$ but actually – from visual observation – should belong to the plateau segment. To correct this, a segment cardinality (represents the duration of each segments) comparison was performed, thus any collection of wave samples with a cardinality less than $T_{min}$ would not be considered as a segment, but combined with the previous segment.

The peak detection scheme consists of local extrema determination outside the resting potential segment. The depolarisation segment begins on the end of each resting potential segment, and ends on the first extremum on each non-resting potential segment. The repolarisation segment begins on the last extremum on each non-resting potential segment, and ends on the start of each next resting potential segment. The region between the depolarisation segment and the repolarisation segment classified as the plateau segment.

A union of one resting potential segment, one depolarisation segment, one plateau segment, and one repolarisation segment defined as one cycle of EGG waveform.

## Electrogastrogram Classification Algorithm

Visually there are at least three types of EGG profiles. From 72 EGG recordings obtained, 63 of them consisted of the resting potential – depolarisation – plateau – repolarisation segments with absolute average resting potential value less than 1 mV, we define this EGG profile as type-λ EGG. Few of the EGG (3 out of 72) observed to be dominantly consisted of the resting potential segment, with resting potential value less than 0 V (ranging from -40 mV to -130 mV), we define this EGG profile as type-ε EGG. The rest of them (6 out of 72), consisted of the resting potential – depolarisation – plateau – repolarisation segments with absolute resting potential value more than 10 mV, we define this EGG profile as type-ρ EGG.

Every electrogastrogram that met (1), and (2) classified as type-λ. Type-ε EGG given for all electrogastrograms those did not fulfill both (1), and (2). Whereas type-ρ was for any other EGG recordings that only met (2).

$$[V_r] < N/c_{min} \qquad (1)$$
$$c > c_{min} \qquad (2)$$

The value of the minimal cycle count ($c_{min}$) used was 3 for recording duration of 6 minutes and 40 seconds. $[V_r]$ is the cardinality of the resting segment for 10 Hz sample rate, where $c$ is the number of cycles in one EGG recording. $N$ is the total cardinality of one EGG recording. This algorithm proved to be adequate even without the resting potential value consideration.